\documentclass[twocolumn,showpacs,preprintnumbers,amsmath,amssymb]{revtex4}

\usepackage{graphicx}
\usepackage{dcolumn}
\usepackage{bm}
\usepackage[rflt]{floatflt}
\usepackage{bibtopic}

\begin{document}


\title {Interplay between the magnetic anisotropy contributions of Cobalt nanowires}

\author{J. S\'anchez-Barriga$^{1 , 2, *}$, M. Lucas$^3$, F. Radu$^2$, E. Martin$^1$, M. Multigner$^{4 , 5}$, P. Marin$^1$, A. Hernando$^1$ and G. Rivero$^1$}

\address{$^1$Instituto de Magnetismo Aplicado (UCM-ADIF-CSIC), P.O.Box 155, 28230, Las Rozas, Madrid, Spain\\
$^2$Helmhotz-Zentrum Berlin f\"{u}r Materialien und Energie, Albert-Einstein-Strasse 15, D-12489 Berlin, Germany\\
$^3$Technische Universit\"{a}t Berlin, Institut f\"{u}r Theoretische Physik, Hardenbergstr. 36, D-10623 Berlin, Germany\\
$^4$Centro Nacional de Investigaciones Metal\'urgicas (CENIM-CSIC), Avd. Gregorio del Amo 8, 28040, Madrid, Spain\\
$^5$Centro de Investigaci\'on Biom\'edica en Red en Bioingenier\'ia, Biomateriales y Nanomedicina (CIBER-BBN), Madrid, Spain}
\date{\today}

\begin{abstract}

 We report on the magnetic properties and the
crystallographic structure of the cobalt nanowire arrays as a
function of their nanoscale dimensions. X-ray diffraction
measurements show the appearance of an in-plane HCP-Co phase
for nanowires with 50 nm diameter, suggesting a partial
reorientation of the magnetocrystalline anisotropy axis along the
membrane plane with increasing pore diameter. No significant changes in the magnetic behavior of the nanowire system are observed with decreasing temperature, indicating that the effective magnetoelastic anisotropy does not play a dominant role in the remagnetization processes of individual nanowires. An enhancement of
the total magnetic anisotropy is found at room temperature with a
decreasing nanowire diameter-to-length ratio (d/L), a
result that is quantitatively analyzed on the basis of a
simplified shape anisotropy model.

\end{abstract}
 \maketitle

The discovery of interlayer exchange
coupling~\cite{IEC-Grunberg-1988} and giant magnetoresistance
\cite{GMR-Baibich-1988,GMR-Binasch-1989} have promoted a
tremendous advance of storage media, readout sensors, and
magnetic random access memory (MRAM). By further reducing the
lateral size of the magnetic structures an increased performance
of the devices is achieved together with a sustained interest
for fundamental understanding of low dimensional magnetism. For
antiferromagnetic materials the finite-size effects lead to a
scaling of the intrinsic magnetic properties like the ordering
temperature and anisotropy as-well as to an enhanced contribution
of the magnetically disordered surface to the macroscopic
magnetization~\cite{morup:2007:jpcm,salabas-2006}. For
ferromagnetic materials with reduced dimensions, the presence of a
finite magnetization leads to an increased shape anisotropy which
depends strongly on the geometry of the
objects~\cite{martin-2003}. This manifests stronger when the
lateral dimensions are touching to the nanoscale regime. Besides
the enhanced shape anisotropies, the remagnetization processes
like coherent rotation and curling modes are more favorable
against the domain wall movements. Ultimately, the manipulation of
the dynamical magnetic properties of such structures in ultrashort
time scales \cite{Parkin-Science-2008} needs to be complementary
to an appropriate characterization and control of their magnetic
and geometrical properties.

In recent years, with the aim of exhaustively tailoring and
controlling properties such as perpendicular magnetic
anisotropy (PMA) \cite{Richter-JPhysD-2007}, a large
variety of geometries for small-sized elements is being
produced by different techniques. Magnetic
nanoparticles \cite{Wernsdorfer-JAP-2000},
nanotubes \cite{Bachman-JAP-2009}, micron-sized
rectangular patelets and dots
\cite{Bolte-JAP-2006}, nanowires \cite{Benitez-PRL-2008} or
ultrathin films are typically fabricated by using various
combinations of state-of-the-art modern experimental tools.
Techniques such as electron beam lithography or imprint
lithography \cite{Smyth-AP-1988}, molecular beam epitaxy (MBE) or
magnetron sputtering are combined to produce high-quality magnetic
structures of ultrasmall sizes~\cite{martin-2003}. Among them,
lithography represents a top-down fabrication technique where a
bulk material is reduced in size to a nanoscale pattern.
Alternatively, electrodeposition \cite{Shiraki-IEEE-1985} is a
very simple and inexpensive bottom-up technique which allows
fabrication of large arrays of magnetic nanowires. Combined with
the use of self-assembling methods for the deposition of membranes
it has the particular advantage of producing systems with high PMA
and ultra-high densities. In order to control the anisotropy of
such a system, it is important to understand how the magnetic
properties depend on the fabrication parameters, and thus on the
geometrical and structural properties of the nanowires. Changes in
the electrodeposition parameters like chemical composition,
temperature and pH of the electrolyte \cite{Darques-JAPD-2004},
deposition time or electrodeposition voltage
\cite{Ganesh-ASC-2005} will result in arrays of nanowires with
different intrinsic magnetic properties.

 We focus our study on the influences of the reduced
diameter-to-length ratio (d/L) on the magnetic effective
anisotropies. Our main finding is that the
otherwise complex demagnetization treatment of the shape
anisotropy can be reduced to a much simpler expression for the
coercive fields of nanoscale Co arrays, which is probed
experimentally. The Co nanowires with lengths between 1
and 6 $\mu$m have been electrodeposided into the pores
of track-etched polycarbonate membranes with nominal
pore diameters of 30 and 50 nm. Previous to
electrodeposition, the membranes were covered by a 100 nm Cu film
by sputtering technique. This acts as an electrode during the
fabrication process. An agitation system was used in order to
avoid hydrogen evolution over the sample and to obtain a
homogeneous growth inside the nanopores. Cobalt nanowires were
grown in a mixed solution \cite{Whitney-Science-1993} (pH=4.5) of
CoSO$_4\cdot$7H$_2$O (252 g/L), H$_3$BO$_3$ (50g/L) and NaCl (7 g/L)
at 25 $\textordmasculineº$C with a constant applied voltage of
-0.95 V. A saturated calomel electrode was used as reference in
potentiostatic mode, and a Co film as counter electrode. By
measuring the experimental development of the deposition current
in the electrochemical cell, we interrupted the growth process at
different times before the complete filling of the nanopores. This
allows a reliable control of the average nanowires length. The
crystallographic structure and morphology of the samples were
investigated by x-ray diffraction and scanning electron microscope
(SEM). The field dependent magnetization hysteresis of the
nanowires was measured at room temperature in a LDJ Vibrating
Sample Magnetometer (VSM) with a external magnetic field of 1 T
applied parallel (H$^{||}$) and perpendicular (H$^{\perp}$) to the
nanowire symmetry axis. Temperature dependent measurements were performed
in a Quantum Design XL7 superconducting quantum interference device magnetometer in the (5-300) K range.

\begin{figure}[tb]
\vspace{0.3cm}
\centering
\includegraphics [width=0.39\textwidth]{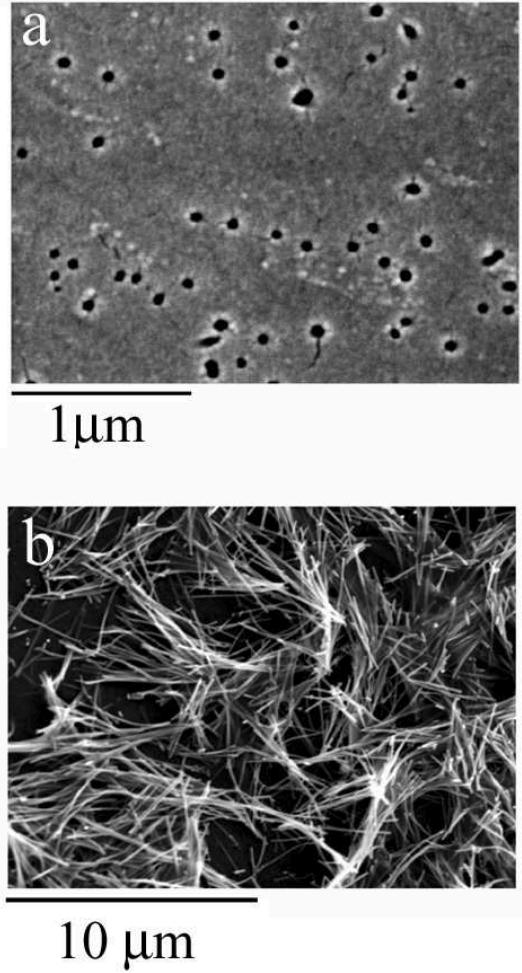}
\caption{SEM images of (a) a small area of a track-etched polycarbonate membrane and (b) Co nanowires after dissolution of the substrate. The images were recorded at 24KV and 43000X and 6000X respectively.}
\label{fig:Fig1}
\end{figure}

Figures~\ref{fig:Fig1}a and~\ref{fig:Fig1}b show the high
resolution SEM micrographs of the membrane surface and nanowires
morphology, respectively. Both measurements were performed with a
JEOL Scanning Microscope JSM-6400 working at 24 kV.
Figure~\ref{fig:Fig1}a corresponds to a polycarbonate membrane
with 6 $\mu$m thickness and 50 nm nominal pore diameter. Some
defects are observed, probably due to stressing effects, as well
as a low ordering degree of nanopores. The pore density and pore
diameter distributions can be estimated from this type of images,
as it has been described in details
elsewhere~\cite{Sanchez-Barriga-JMMM-2007}. The measured mean pore
density of (6.1 $\pm$ 2.4)$\cdot$10$^8$ nanopores$\cdot$cm$^{-2}$
is in good agreement with the value of 6$\cdot$10$^8$
nanopores$\cdot$cm$^{-2}$ given by the manufacturer.
Figure~\ref{fig:Fig1}b shows Co nanowires with 50 nm nominal
diameter and $\sim$ 3 $\mu$m length after dissolution of the
polycarbonate membrane. While inside the membrane all nanowires
are parallel to each other, after its dissolution they reorganize
in different positions all over the sample area. In this way, the
length of the nanowires can be measured more accurately and
compared to the chronoamperometric measurements as explained in
Ref.~\cite{Sanchez-Barriga-JMMM-2007}. Note that in the
Fig.~\ref{fig:Fig1}a, the apparent deviations from a circular
shape of the nanowires occurs only near the surface. This is due
to the fabrication process of the polycarbonate membranes which
typically leads to a conical opening of the nanopores which
extends a few nanometers deep from the membrane surface.
Therefore, we have taken particular care and ended the
deposition process before the nanopores were completely filled
with the magnetic material.

\begin{figure}[th]
\centering
\includegraphics [width=0.45\textwidth]{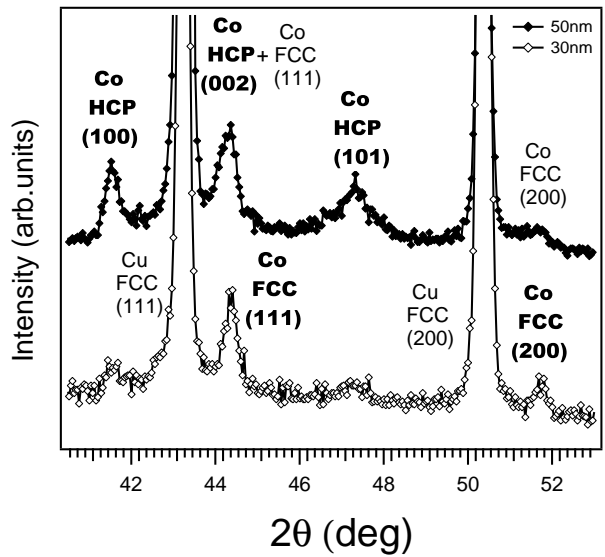}
\caption{Comparison of the x-ray diffraction patterns of electrodeposited Co nanowires with 30 nm and 50 nm nominal diameter.}
\label{fig:Fig2}
\end{figure}

For the x-ray diffraction (XRD) measurements we have utilized a
SIEMENS D-5000 diffractometer providing a Cu-K$_\alpha$ radiation
($\lambda=1.5418 {\AA}$) under working conditions of 40 kV and 30
mA. The measurements were done prior to dissolution of the
polycarbonate membrane (as seen in Fig.~\ref{fig:Fig1}a), when all
nanowires are parallel to each other and the sputtered Cu layer on
one side of the membrane was still present. Figure~\ref{fig:Fig2}
shows the $\theta-2\theta$ diffraction patterns of Co nanowires
with $\sim$6 $\mu$m length and nominal pore diameters of 30~nm and
50~nm. Contributions from the Cu substrate appear as intense peaks
corresponding to Cu-FCC (111) and (200) orientations. The two
diffractograms exhibit different crystalline structures as a
function of the nanowire diameter. The diffraction pattern of
nanowires with 30~nm diameter is dominated by a highly textured
Co-FCC phase oriented along the [111] and [200] directions. The
nanowires with 50~nm diameter exhibit, however, a more complicated
diffraction pattern which indicates that a mixture of Co-HCP and
Co-FCC phases is present. This suggests that Co nanowires are
composed of hcp and fcc crystalline Co
segments~\cite{Zuxin-JAP-2009}.
 We basically observe two new prominent
structures corresponding to a Co-HCP phase oriented along the
[100] and [101] directions. Their coexistence gives rise to an
asymmetric peak appearing near the expected angular positions of
the Co-HCP (002) and Co-FCC (111) phases. The double structure
behind this peak cannot be resolved in this case, indicating that
the degree of coexistence between the two crystalline phases is
dominated by the Co-HCP contribution.

In addition, other features, as the one corresponding to Co-FCC
(200) phase, are barely distinguishable from the background
signal. Since the final texture of the nanowires also depends on
the plating procedure, it is still unclear why the amount of FCC
phase increases with decreasing nanowire diameter. In recent
studies, a coexistence of two different Co phases with different
orientations for nanowires with 200 nm in diameter was also found
\cite{Zuxin-JAP-2009}. A transition to a Co-HCP crystalline
structure was shown for nanowires with 65 nm in diameter and a pH
and diameter-dependent phase diagram for electroplated Co
nanowires was proposed. Our results give complementary information
to the proposed phase diagram in Ref.~\cite{Zuxin-JAP-2009},
indicating that by further reduction of the nanowire diameter a
second transition to a Co-FCC phase occurs for nanowire diameters
of $\sim$30 nm. Other investigations of Co nanoparticles
\cite{Kitakami-PRB-1997} show that the presence of a HCP phase or
a FCC phase can be related to the particle diameter. In that case,
the transition from HCP to FCC was ascribed to a size effect due
to the lower surface energy of the FCC phase.

\begin{figure*}[th]
\centering
\includegraphics [width=1\textwidth]{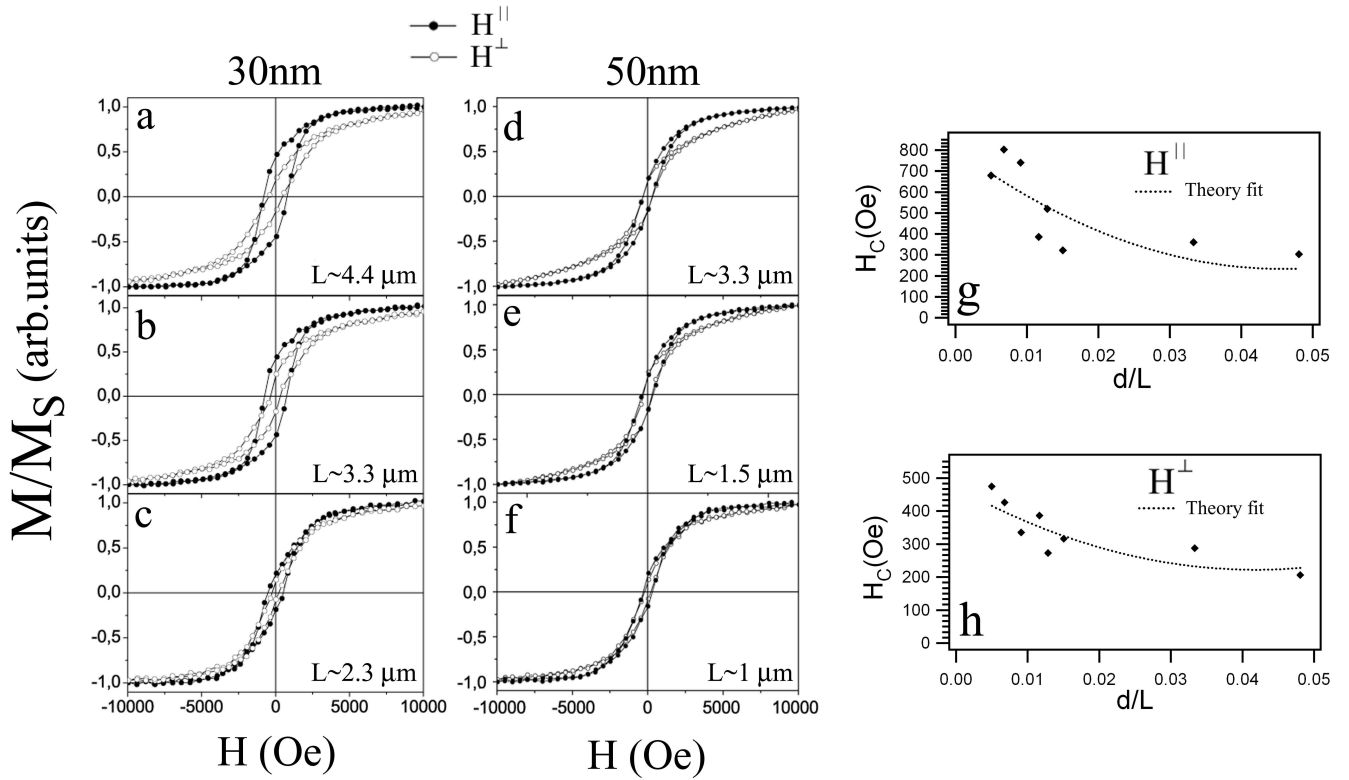}
\caption{Room temperature field dependent magnetization hysteresis of Co nanowires with different lengths and diameters. Figs. 3a, 3b and 3c: Nanowires with 30 nm in diameter and lengths of $\sim$4.4 $\mu$m, $\sim$3.3 $\mu$m, $\sim$2.3 $\mu$m, respectively; figs. 3d, 3e and 3f: Nanowires with 50 nm in diameter and $\sim$3.3 $\mu$m,$\sim$ 1.5 $\mu$m, $\sim$1 $\mu$m in length. Figs 3g and 3h: General dependence of the coercive field as a function of the ratio (d/L) of the nanowires when the magnetic field is applied parallel (H$^{||}$) and perpendicular (H$^{\perp}$) to the nanowire axis.}
\label{fig:Fig3}
\end{figure*}

Figures~\ref{fig:Fig3}a-f show the resulting hysteresis loops
measured for nanowire arrays with different diameters and lengths.
Figures~\ref{fig:Fig3}g-h show the extracted coercive field values
and the corresponding theoretical fits (see below) as a function of the nanowire diameter-to-length ratio for both (H$^{||}$) and (H$^{\perp}$) externally applied fields, respectively. A better understanding of the observed dependence of the coercive field with the nanoscale dimensions observed in figures~\ref{fig:Fig3}g-h can only be achieved by analyzing the effects of the different anisotropy contributions to the total anisotropy of the system, as it will be shown in the following. The first observation is a pronounced reduction of the coercivity and remanent magnetization values for nanowires with 50 nm in diameter (Figs.~\ref{fig:Fig3}d-f). Besides, by comparing the axial and
transverse hysteresis loops for each case, we observe how the
effective magnetic anisotropy along the nanowire axis increases
with increasing length for both diameters. The longer the
nanowires, the larger the remanent magnetization and coercive
field values for an applied field parallel to the nanowire axis
($H^{||}$). This effect appears more pronounced for nanowires with
30 nm in diameter (Figs.~\ref{fig:Fig3}a-c).

We argue below that the effective magnetic anisotropy is mainly
oriented along the nanowire axis in all cases, and that the shape
anisotropy of the nanowire arrays is playing a more dominant role
than the dipolar interaction for our particular system. The shape
anisotropy of an infinitely long wire would result in a
squared hysteresis loop exhibiting a finite coercive field for an
applied field along the wire H$^{||}$ and zero coercive field for
a perpendicular applied field to the wire H$^{\perp}$. Within the
Stoner-Wohlfarth model, a particular case is the so called
prolate spheroid geometry of magnetic objects.
If the ellipsoid is uniformly magnetized, the magnetic field
inside the ellipsoid depends directly on the demagnetization
tensor ($\vec{H}_{in}{\sim}-N\vec{M}$). Since the trace of the
tensor N must be 1, for a limiting case of a infinitely long
cylinder in which the shape anisotropy plays the dominant role one
would obtain $N_x=N_y=1/2$ and $N_z=0$. In a general case, the
demagnetizing factor along the long axis of the ellipsoid (or, by
analogy, along the cylinder ) depends on second order terms of the
type $(d/L)^2$ and it approaches 0 when L$\rightarrow\infty$.
Since $\vec{H}_{c}{\sim}\vec{H_a}-N\vec{M}$, this leads to the
following expectation for the coercive field:
$\vec{H}_{c}\rightarrow\vec{H_a}$ when N$\rightarrow$ 0, where
$H_a$ is the intrinsic anisotropy field. Therefore, a system with
finite dimensions which is dominated by shape anisotropy
($d{\ll}L$) exhibits a finite $N_z$ ($N_z\neq 0$). Our
observations in Fig. 3, where coercive field values for H$^{||}$
are larger than the for H$^{\perp}$, are fully consistent with
shape anisotropy origin of the enhanced coercivity as discussed
qualitatively within the Stoner-Wohlfart model. Nevertheless,
other contributions to the effective anisotropy of the whole
nanowire array, other than the shape induced anisotropy, may
influence the observed magnetic behavior: dipolar interactions among nanowires, magnetocrystalline anisotropy and magnetoelastic
anisotropy \cite{Escrig-PRB-2007}. As it has been shown for nanowires grown
in alumina membranes, the former certainly depends strongly on the
inter-wire distances and typically manifests through a decreasing
coercive field along H$^{||}$ with an increasing ratio of nanowire
diameter to inter-wire spacing \cite{Vazquez-JMM-2005}. Although
the use of randomly distributed array of nanopores which is
characteristic of the polycarbonate membranes does not allow an
exhaustive control of the inter-wire distances, an average value
of $\sim$500 nm can be deduced, which is one order of magnitude
larger than the typical nanowire diameter. Therefore, for our case the dipolar
interactions have a minor effect for sufficiently long nanowires.
A reminiscent influence of the dipolar interaction on the magnetic
behavior would manifest through a small contribution to the
reduced remanent magnetization for shorter wires. Further
discussion about the role of the magnetocrystalline anisotropy is
provided below by interpreting the measured hysteresis loops in
combination with the different orientations of the crystalline
structure observed in the x-ray diffraction patterns of Fig. 2.

For nanowires which are 30 nm in diameter, the [111] direction of
the Co-FCC phase is mainly oriented along the axis of the
nanowires, with the basal (111) planes being parallel to the
membrane plane. For nanowires which are 50~nm in diameter, the
Co-HCP [100] and [002] directions are oriented perpendicular and
parallel to the nanowire axis, respectively. Moreover, the c-axis
of the HCP (101) phase makes an angle of 32$\textordmasculineº$
with respect to the nanowire axis.
As a result, the effective crystalline alignment resulting from
the different Co-HCP orientations observed in the diffraction pattern
would lead to a reinforcement of the total magnetocrystalline anisotropy of the
system along the H$^{\perp}$ direction. For a Co-FCC(111) phase
oriented along the nanowire axis, the magnetocrystalline
anisotropy energy density K$_1$=6.3$\cdot$10$^5$ erg/cm$^3$ is
almost one order of magnitude smaller than the shape anisotropy
energy density K$_S$=$\pi\cdot$M$^2_S$=6.0$\cdot$10$^6$
erg/cm$^3$, whereas for a Co-HCP structure, they are of the same
order of magnitude (K$_1$=5$\cdot$10$^6$ erg/cm$^3$
$\sim$K$_S$=$\pi\cdot$M$^2_S$=6.0$\cdot$10$^6$ erg/cm$^3$)
\cite{Li-JPCM-2004}. Hence, for nanowires with 50 nm diameter, we
can expect a more pronounced interplay of both magnetocrystalline
and shape anisotropies that results in a slightly weaker effective
anisotropy along the nanowire axis. This leads to a further
reduction of the coercive field and remanent magnetization for
H$^{||}$ as compared to the 30 nm wires. In this respect, the
reduced coercivity for larger diameters depends not only on the
geometrical properties, i.e, the nanowire diameter dependence of
the coercive field \cite{Gao-PRB-2007}, but also on the partial
influence of the crystalline orientation of the Co-HCP phase, i.
e., on magnetocrystalline anisotropy.

\begin{figure*}[th]
\centering
\includegraphics [width=0.83\textwidth]{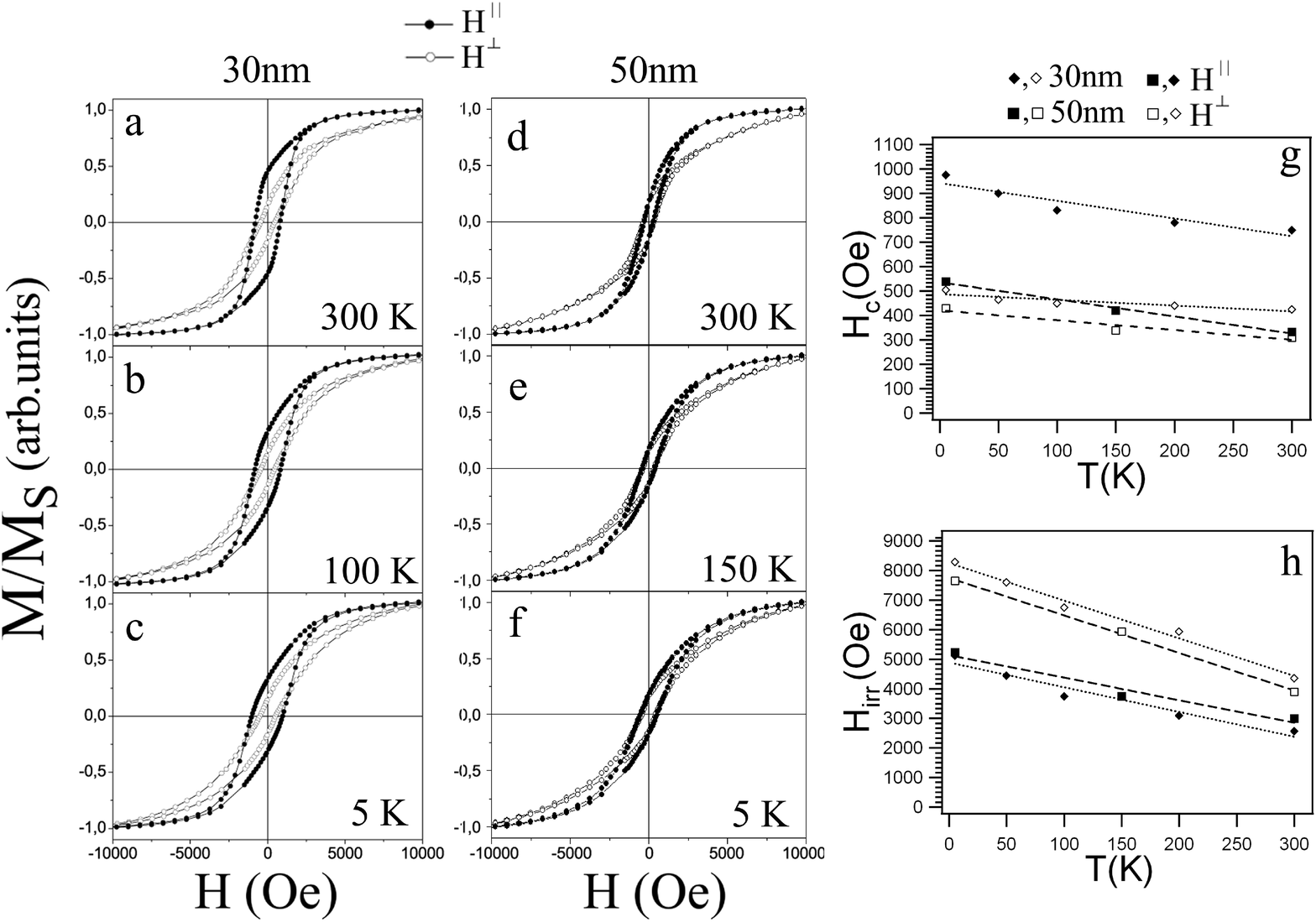}
\caption{Field dependent magnetization hysteresis of Co nanowires at different temperatures. Figs. 3a, 3b and 3c: Nanowires with 30 nm in diameter and $\sim$4 $\mu$m in length; figs. 3d, 3e and 3f: Nanowires with 50 nm in diameter and $\sim$3 $\mu$m in length. Figs 3g and 3h: General dependence of the coercive and irreversible fields as a function of temperature for both parallel (H$^{||}$) and perpendicular (H$^{\perp}$) appied magnetic fields. Dotted and dashed lines result from linear fits of the experimental data for nanowires with nominal diameters of 30 and 50 nm respectively.}
\label{fig:Fig4}
\end{figure*}

As to further investigate the validity of the aforementioned
effects, i.e, the interplay between magnetocrystalline and shape
anisotropies, we present in Fig.~\ref{fig:Fig4} temperature
dependent measurements for nanowires with 30 and 50 nm diameter
and $\sim$4 and $\sim$3 $\mu$m in length, respectively. Such type
of measurements allow us to determine if other contributions,
particularly the ones arising from magnetoelastic effects may
influence the magnetic behavior of the nanowires system at room
temperature. Figures~\ref{fig:Fig4}a-f show the resulting
hysteresis loops measured for both systems of nanowire arrays at
different temperatures. Figures~\ref{fig:Fig4}g-h show the
extracted coercive field ($H_{c}$) and irreversible switching
field ($H_{irr}$) values as a function of temperature for both
parallel (H$^{||}$) and perpendicular (H$^{\perp}$) applied
magnetic fields. $H_{irr}$ is the field where the magnetization
changes irreversibly (i.e, where the hysteresis opens). In
Figures~\ref{fig:Fig4}a-f we do not observe significant changes in
the hysteresis loops when the temperature decreases down to 5 K.
The main observation is a quasi-monotonic decrease of $H_{c}$ and
$H_{irr}$ with increasing temperature, as summarized in
figures~\ref{fig:Fig4}g-h. This effect which is observed for both
applied field directions, appears to be less pronounced in the
case of the $H_{c}$ values extracted for H$^{\perp}$. As shown in
figures~\ref{fig:Fig4}g-h, we performed linear fits to the
experimental data from which we estimate the relative changes in
the angle of orientation of the anisotropy axis ($\Theta$),
$H_{c}$ and $H_{irr}$. We obtain
$\Theta\sim$35$\textordmasculineº$ and a partial reorientation of
$\sim$6$\textordmasculineº$ towards the nanowire axis when the
temperature is decreased. For H$^{||}$, the changes of $H_{c}$
with respect to its low temperature value are $\sim$23\% and
$\sim$38\% for nanowires with 30 and 50 nm in diameter,
respectively. For an applied field H$^{\perp}$, these changes are
$\sim$15\% and $\sim$24\%, respectively. The relative increase in
$H_{irr}$ is $\sim$47\% in all cases. Note that in general, (i)
this effect is more pronounced in the $H_{irr}$ values and that
(ii) $H_{c}$ is larger for H$^{||}$ than for H$^{\perp}$, whereas
$H_{irr}$ is smaller for H$^{||}$ than for H$^{\perp}$. This
differences are expected from the Sthoner-Wohlfarth theory, and
they can be qualitatively explained by considering the calculated
azimuthal dependence of both $H_{irr}$ and $H_{c}$ for a magnetic
system with uniaxial anisotropy \cite{Radu-TMP-2008}. Our results
indicate that only a moderate increase of the effective anisotropy
of the nanowire system along the nanowire axis occurs at low
temperatures, resulting in a rather small reorientation of the
anisotropy axis along the nanowires. Previous studies of Ni
nanowires showed that magnetoelastic effects can have profound
influences on the magnetic properties of the arrays
\cite{Dubois-PRB-2000,Kumar-PRB-2006}, typically leading to a
reorientation of the magnetic easy axis of the nanowire system
from the parallel to perpendicular configuration with decreasing
temperature. At low temperatures, the nanowires are under stress
due to the large mismatch between the thermal expansion
coefficients of Co ($\alpha_{Co}\sim$ 13$\cdot$10$^{-6}$ K$^{-1}$)
and the polycarbonate membrane ($\alpha_{Polyc}\sim$
67$\cdot$10$^{-6}$ K$^{-1}$). Since $\alpha_{Polyc}$ \textgreater
$\alpha_{Co}$, the polycarbonate template tends to contract more
than the Co during cooling, resulting in a transverse compressive
force which acts perpendicular to the nanowire axis and leads to
an expansion of the nanowires along their axis. The axial strain
produced in this process can be calculated as
$\epsilon_{||}$=-$\nu\Delta$T($\alpha_{Co}$-$\alpha_{Polyc}$),
where $\nu$ is the Poisson\textquoteright s ratio of Co
($\nu_{Co}$=0.31) and $\Delta$T=295 K when the temperature varies
from 300 to 5 K. This is equivalent to an axial tensile stress of
$\sigma_{||}$=E$_{Co}\epsilon_{||}\sim$1.03 GPa, where
E$_{Co}\sim$209 GPa is the Young\textquoteright s modulus of Co.
Considering that the saturation magnetostriction constant of Co is
negative ($\lambda^{Co}_{s}\sim$-55$\cdot$10$^{-6}$)
\cite{Kazakova-PRB-2006,Pirota-PRB-2007}, such an axial effective
tensile stress will decrease the effective anisotropy along the
nanowire axis and, therefore the $H_{c}$ and $H_{irr}$ values, in
contrast to our observations. In some instances, when the
magnetoelastic effects are very strong \cite{Pirota-PRB-2007}, the
interplay between magnetoelastic and shape anisotropies may lead
to a reorientation of the anisotropy axis at certain crossover
temperature below which $H_{c}^{\perp}$ \textgreater $H_{c}^{||}$.
The maximum contribution of the transverse magnetoelastic
anisotropy energy density can be estimated as
K$_{me}$=3$\lambda^{Co}_{s}\sigma_{||}$/2$\sim$-8.4$\cdot$10$^{5}$erg/cm$^3$
which is one order of magnitude smaller than the shape anisotropy
energy density K$_S$=6.0$\cdot$10$^6$erg/cm$^3$. Therefore, the
effective magnetoelastic anisotropy is not playing a dominant role
in the remagnetization processes of individual Co nanowires. 
Note that the calculated value of K$_{me}$ represents an upper
limit since we assume perfect adhesion of the nanowires to the
pore walls. In this simplified calculation, we have neglected (i)
the weak dependence of the E$_{Co}$ with the nanowire length, (ii)
the losses in the compression energy due to the large mismatch
between the Young\textquoteright s modulus of Co and the
polycarbonate membrane (E$_{Polyc}\sim 2GPa$\textless\textless
E$_{Co}$),(iii) nonlinear effects in the temperature dependent
expansion coefficients and (iv) other second order effects in the
temperature dependence of the elastic properties. We attribute the
observed effects to a preferential increase of the effective
magnetocrystalline anisotropy
 with
decreasing temperature. While cooling down to 5 K, the $H_{c}$ and
$H_{irr}$ are increasing due to the thermal energy which acts
against the anisotropy energy.
At low temperatures, the relative changes in $H_{c}$ and $H_{irr}$
will depend on the size and temperature dependent behavior of the
magnetocrystalline anisotropy, and therefore on the orientation of
the different crystalline structures appearing in the x-ray
diffraction patterns of Fig. 2. For nanowires with 30 nm in
diameter, the relative increase of the magnetocrystalline
anisotropy along the nanowire axis is $\sim$27\%, whereas
$\sim$40\% for nanowires with 50 nm in diameter. This indicates
that the dominant contributions in the diffraction patterns are
the Co-HCP(002) and Co-FCC(111) phases for nanowires with 50 and
30 nm in diameter, respectively.

In order to gain more insight on the interplay between both
magnetocrystalline and shape anisotropies, we would like to come back to the results shown in Figs. 3g and 3h and provide further theoretical analysis of the observed room temperature dependence of the coercive field as a function of the
diameter-to-length ratio (d/L) of the nanowires. Such type of analysis gives more information about the origin of the dominant anisotropy contributions to the total magnetic anisotropy
of the system and, particularly, how strongly the remagnetization
process of the nanowires depend on the shape anisotropy, i.e, on
the nanowire dimensions. Based on the aforementioned argumentations, our theoretical analysis does not consider contributions to the effective anisotropy of the nanowire system arising from magnetoelastic effects and dipolar interactions among nanowires. As previously mentioned, the results of Figs. 3g and 3h are
presented together with a fit function
$H_c^{||,\perp}=f^{||,\perp}(d/L)$ which explicitly shows a
size-dependent behavior as it will be deduced in the following.
The physical origin of this dependence are demagnetizing fields
\cite{Beleggia-JPDAP-2005,Leven-PRB-2005}, which give rise to the
shape anisotropy of the nanowires and, therefore, it allow us to
check the validity of a simple model in which the shape anisotropy
is the dominant contribution to the observed size-dependent
coercive field behavior. The relation between the demagnetizing
factors N$^{||}$ or N$^{\perp}$ and the structure dimensions can
be expressed in terms of the hypergeometric Gauss
function${}^{\phantom{1}}_{\phantom{1}2}\text{F}_{1}$
\cite{Millev-JPhysD-2003}. In the limit in which $d{\ll}L$ and
taking into account that any given finite circular cylinder with
diameter d and length L possesses rotational symmetry about its
geometrical axis, for $\kappa=d/L$, N$^{||}$ can be written as
\cite{Beleggia-JPhysD-2006}:
\begin{equation}
N^{||}=\frac{4}{3\pi}\kappa-8\kappa^2+ O(\kappa^4)
\end{equation}
which represents, a second order dependence of the demagnetizing
factor on the diameter-to-length ratio of the nanowires. We can
consider that the expressions for both coercive fields
H$_{c}$$^{||}$ and H$_{c}$$^{\perp}$ are given by
\cite{Radu-TMP-2008}:
\begin{equation}
H_{c}^{||}=\frac{2K_{eff}}{\mu_0M_s}|cos\Theta|; \quad H_{c}^{\perp}=\frac{2K_{eff}}{\mu_0M_s}|sin\Theta|
\end{equation}
with $K_{eff}$ being the effective anisotropy constant
of the ferromagnet, $M_s$ the saturation magnetization,
and $\Theta$ the angle between the applied field and
the anisotropy axis of the system. Considering
that $N^{||}-N^{\perp}=(3N^{||}-1)/2$, a widely
used phenomenological expression for $K_{eff}$ is given by:
\begin{align}
& \epsilon_{k}K_M-\frac{(1-3N^{||})\mu_0M_s{^2}}{4} \quad \text{if } d<d_{c}\\
& \epsilon_{k}K_M+\frac{c(N^{||})A}{(d/2)^2}-\frac{N^{||}\mu_0M_s{^2}}{2} \quad \text{if } d>d_{c}
\end{align}
where $\epsilon_{k}$ is the real-structure-dependent
Kronm\"{u}ller parameter \cite{Kronmuller-PSSb-1987}, $K_M$ is
the effective magnetocrystalline anisotropy constant, d$_{c}$
is the critical coherence diameter and A is the exchange
stiffness constant (A=30$\cdot$10$^{-12}$J/m for Co).
If d$<$d$_{c}$, then the dominant exchange interaction
yields coherent (uniform) rotation (equation (3)),
whereas if d$>$d$_{c}$ magnetostatic interactions give
rise to curling reversal (equation (4)). The critical
diameter, defined as d$_{c}$=3.68$\sqrt{A/\pi M_s{^2}}$
yields to d$_{c}\sim$32 nm for Co. The second term on
the right hand of equation (3) represents the shape
anisotropy contribution to the effective anisotropy
of the system, while in equation (4) the exchange
term partly compensates for the absence of shape
anisotropy in a proper sense. Considering that
$c(N^{||})=(M_sd_c){^2}N^{\perp}$ and by substituting
(3) or (4) into (2) and then introducing (1), after
some manipulation the coercive field becomes:
\begin{equation}
H_{c}^{||,\perp}=\alpha^{||,\perp}-\beta^{||,\perp}\left(\frac{d}{L}\right)+\gamma^{||,\perp}\left(\frac{d}{L}\right)^2
\end{equation}
The relation between the physical constants and the fitted
parameters $\alpha^{||,\perp}$, $\beta^{||,\perp}$ and
$\gamma^{||,\perp}$ can be written as follows:
\begin{equation*}
\frac{(\alpha^{||},\beta^{||},\gamma^{||})}{|cos\Theta|}=\frac{(\alpha^{\perp},\beta^{\perp},\gamma^{\perp})}{|sin\Theta|}=(A,B,C)
\end{equation*}
with A, B and C given by:
\begin{align*}
& A=\alpha\left(\frac{2\epsilon_{k}K_M}{\mu_0M_s}+\frac{g M_s}{2}\right);B=\frac{4\beta M_s}{3\pi}\left(1+\frac{g}{2}\right); \\
& C=8\gamma M_s\left(1+\frac{g}{2}\right)\quad\text{with}\quad\ g=\left(\frac{d_c}{d}\right)^2
\end{align*}
where g=1 for coherent rotation. We have introduced the parameters
$\alpha$, $\beta$ and $\gamma$ in order to account for
quantitative deviations from the situation in which the
magnetization reversal processes are driven by coherent rotation
or curling mechanisms. Certainly, since the diameters of the
fabricated nanowires are near to the limit for coherent rotation
it is not straightforward to deduce which magnetization reversal
process is the preferred mechanism. For curling processes and within the present approximation, we assume that $d_{c}$ is not decreasing strongly with increasing
d/L. That means that $0.20\leq g/2\leq 0.56$ for nanowires with 50
and 30 nm respectively. The results of the fit using the function
(5) show qualitative agreement to the experiment and are
represented in Figs. 3g and 3h. The fitted parameters are given in
Table I. From the experimental data obtained at maximum and
minimum d/L values we estimate a nearly constant angle of rotation
respect to the nanowire axis of
$\Theta\sim$32$\textordmasculineº$, and its variation is below
8$\textordmasculineº$ for a decreasing structure size. In
agreement to it, the theoretical fit gives
$\Theta$=(28$\pm$3)$\textordmasculineº$. Therefore, we suggest
that the observed changes in the coercive field are mainly due to
an increase of the effective magnetic anisotropy $K_{eff}$ of the
system, rather than due to a rotation of the anisotropy axis
towards the nanowire axis. Considering $\epsilon_{k}$=1 and
introducing M$_s$(Co)$\sim$1424Oe, for $\alpha$=1 and g=1 we
estimated K$_M$ $\sim$(1.8$\pm$ 0.4)$\cdot$ 10$^6$ erg/cm$^3$, in
agreement with the average value of $\sim$2.8$\cdot$10$^6$
erg/cm$^3$ obtained from the magnetocrystalline anisotropy energy
densities of Co-FCC and HCP crystalline structures. The estimated value of K$_M$ indicates that the changes in the magnetocrystalline anisotropy density of the system are mostly due
to a mixture of both FCC and HCP crystal orientations, in
agreement with the x-ray diffraction patterns of Fig.2.
\begin{table}[ht]
\centering
\begin{tabular}{c|c|c|c}
\hline\hline
& $\alpha^{||,\perp}$ (Oe) & $\beta^{||,\perp}$ (Oe) & $\gamma^{||,\perp}$ (Oe) \\
\hline
H$^{||}$ & 805$\pm$ 108 & (2.5$\pm$ 0.6)$\cdot$ 10$^4$ & (2.7$\pm$ 1.3)$\cdot$ 10$^5$ \\
H$^{\perp}$ & 472$\pm$ 60 & (1.2$\pm$ 0.3)$\cdot$ 10$^4$ & (1.4$\pm$ 0.6)$\cdot$ 10$^5$ \\
\hline
\end{tabular}
\caption{Values of the fitted parameters $\alpha^{||,\perp}$, $\beta^{||,\perp}$ and $\gamma^{||,\perp}$ for both applied field directions H$^{||}$ and H$^{\perp}$ respectively.}
\end{table}
For g=1, the deviations from the first and second order terms yield to
$\beta\sim$(30$\pm$8) and $\gamma\sim$(18$\pm$9), whereas for
$0.20\leq g/2\leq 0.56$, $\beta\sim$(34$\pm$10) and
$\gamma\sim$(20$\pm$11). Although a qualitative agreement is
reached in the general dependence of the coercive field with d/L,
this deviations indicate that the Stoner-Wohlfarth approximation
is overestimating the coercive field values by a factor of $\sim$3 for
both applied directions. In this respect, it was shown that at the
wire ends a butterfly-type arrangement of the magnetization exists
which reduces the switching field considerably
\cite{Holz-PSSolidi-1968}. Micromagnetic simulations for Ni
nanowires of 40 nm diameter, also showed that the nucleation field
can be typically reduced by a factor of 2 \cite{Suhl-JAP-1997}. In
addition, in the case of an array of nanowires, collective
demagnetization modes have to be taken into account which lead to
a further decrease of the coercive field. The reduced squareness
observed in the hysteresis loops of Fig. 3 might result from the
existence of polycrystalline and surface-related structural
imperfections along the nanowires, leading to a magnetic
localization of the reversal modes \cite{Zheng-JPhys-2000}. The
coherent-rotation and curling modes are delocalized, which means
that the spatial variations of the magnetization along the
nanowire are small. However, local variations of the magnetization
cost some exchange energy but they may be energetically favorable
from the point of view of a local anisotropy, $K_{eff} (r)$. This
competition leads not only to a reduction of the nucleation field
but also to a localization of the magnetization reversal modes and
to the formation of magnetic domains along the wires. Further
understanding of the real-space character of the magnetization
mechanisms can be achieved by considering cooperative effects
between coherent-rotation and curling modes
\cite{Skomski-JPhys-2001}. Actually this type of processes are of
great importance in advanced technology, because they lead to the
formation of interactive magnetic domains which may improve the
thermal stability of the samples or reduce the storage densities
in the case of extended magnets or ultrathin films.

In conclusion, we have shown that the magnetic properties of the
fabricated Co nanowire arrays are dominated by the shape
anisotropy contributions more than the magnetocrystalline
anisotropy, magnetoelastic effects and dipolar interactions among
nanowires. A partial influence of the different crystallographic
orientations on the magnetic hysteresis loops at room temperature
is deduced by analyzing the x-ray diffraction patterns. Low
temperature measurements indicate that the magnetoelastic
anisotropy is not playing a dominant role in the remagnetization
process of the nanowires. The changes are attributed to a
temperature dependent behavior of the magnetocrystalline
anisotropy. We have developed a phenomenological model on the
basis of the Stoner-Wohlfarth theory which reproduces the
dependence of the coercive field on the nanowire
diameter-to-length aspect ratio at room temperature. The increase
of coercive field as d/L decreases is attributed to an enhancement
of the effective magnetic anisotropy of the system for a nearly
constant angle of orientation respect to the applied magnetic
field. The quantitative analysis of the coercive field within the
proposed model suggests additional contributions to
remagnetization mechanisms, discussed in terms of localization and
cooperative effects of the magnetization reversal modes due to
polycrystalline and surface-related structural imperfections in
the nanowires.

$*$ Electronic adress: sbarriga@bessy.de
\end{document}